\documentclass[a4paper,aps,prl,showpacs,preprintnumbers,twocolumn,nofootinbib]{revtex4}

\usepackage{amsmath,amssymb,graphicx}
\usepackage{color}
\usepackage[normalem]{ulem} 

\newcommand{\be}{\begin{equation}}
\newcommand{\ee}{\end{equation}}
\newcommand{\bea}{\begin{eqnarray}}
\newcommand{\eea}{\end{eqnarray}}
\newcommand{\bean}{\begin{eqnarray*}}
\newcommand{\eean}{\end{eqnarray*}}

\newcommand{\HH}{{\cal H}}

\newcommand{\de}{\delta}
\newcommand{\De}{\Delta}

\newcommand{\La}{\Lambda}

\newcommand{\Om}{\Omega}

\newcommand{\si}{\sigma}

\newcommand{\lsim}{\stackrel{<}{\sim}}

\def\id{{\rm 1\kern -2.5pt I}} 

\definecolor{dgreen}{rgb}{0,0.6,0} 

\begin{document}

\preprint{DESY 13-091}

\title{The value of $H_0$ in the inhomogeneous Universe}

\author{Ido Ben-Dayan$^1$, Ruth Durrer$^2$, Giovanni Marozzi$^2$ and Dominik J.~Schwarz$^3$}

\affiliation{$^1$Deutsches Elektronen-Synchrotron DESY, Theory Group, D-22603 Hamburg, Germany\\
$^{2}$Universit\'e de Gen\`eve, D\'epartement de Physique Th\'eorique and CAP,
24 quai Ernest-Ansermet, CH-1211 Gen\`eve 4, Switzerland\\
$^{3}$Fakult\"at f\"ur Physik, Universit\"at Bielefeld, Postfach 100131, 33501 Bielefeld, Germany}

\date{\today}

\begin{abstract} 
Local measurements of the Hubble expansion rate are affected by structures like galaxy clusters 
or voids. Here we present a
fully relativistic treatment of this effect, studying how clustering modifies the mean distance 
(modulus)-redshift relation and its dispersion in a standard $\La$CDM universe. 
The best estimates of the local expansion rate stem from supernova observations at small 
redshifts ($0.01<z<0.1$). It is interesting to compare these local measurements with global fits to 
data from cosmic microwave background anisotropies. In particular, we argue that cosmic variance 
(i.e.~the effects of the local structure) is of the same order of magnitude as the current observational 
errors and must be taken into account in
local measurements of the Hubble expansion rate.
\end{abstract}


\maketitle

The Hubble constant, $H_0$,
determines the present expansion rate of the Universe.
For most cosmological phenomena a precise knowledge of $H_0$ 
is of utmost importance.
In a perfectly homogeneous and isotropic world $H_0$ is defined globally. 
But the Universe contains structures like galaxy clusters and voids. Thus the local expansion rate,
measured by means of cepheids and supernovae at small redshifts, does not necessarily agree with 
the expansion rate of an isotropic and homogeneous model that is used to describe the Universe 
at the largest scales.

Recent local measurements of the Hubble rate 
\cite{Riess, Freedman:2012ny}  are claimed  to be accurate at the few percent level, 
e.g.~\cite{Riess} finds $H_0=(73.8 \pm 2.4) \,\, {\rm km~s}^{- 1} {\rm Mpc}^{-1}$. 
In the near future, observational techniques will improve further, such that the local value of 
$H_0$ will be determined at 1\% accuracy~\cite{future}, competitive with the current precision of 
indirect measurements of the global $H_0$ 
via the cosmic microwave backgound anisotropies \cite{Planck}.

The observed distance modulus $\mu$ is related to the bolometric flux $\Phi$ and the 
luminosity distance $d_{\rm L}$ by ($\log \equiv \log_{10}$)
\be
\label{mudef}
\mu = -2.5 \log [\Phi/\Phi_{10\ \rm pc}] = 5 \log [d_{\rm L}/(10\ {\rm pc})].
\ee
The relation between the intrinsic luminosity, $L$, the bolometric flux, $\Phi$, and the luminosity distance $d_L$  of a source is
$\Phi = L/{4\pi d_L^2}$.
In a flat $\Lambda$CDM  universe with present matter density parameter $\Om_m$ 
the luminosity distance as a function of redshift $z$ is given by
\be
\label{dLLCDM}
 d_L(z) = \frac{1+z}{H_0/c} \int_0^z\!\! \frac{dz'}{\sqrt{\Om_{m}(1+z')^3 +1-\Om_{m}}}. 
\ee
As long as we consider only small redshifts, $z\le0.1$, the dependence on cosmology is weak,
$d_L(z) \simeq c [z + (1-3\Om_m/4) z^2]/H_0$ and the result varies by about 0.2\% when $\Om_m$  
varies within the $2\si$ error bars determined by  Planck \cite{Planck}. However, neglecting the 
model dependent quadratic term induces an error of nearly 8\% for $z\simeq 0.1$. 

The observed Universe is inhomogeneous and anisotropic on small scales and
the local Hubble rate is 
expected to differ from its global value for two reasons. First, any 
supernova (SN) sample is finite (sample variance) and, second, we observe only one realization of a
random configuration of the local structure (cosmic variance). Thus, even for arbitrarily precise 
measurements of fluxes and redshifts, the local $H_0$ differs from the global $H_0$. 
Sample variance is fully taken into account in the literature, but cosmic variance is usually 
not considered. 

In the context of Newtonian cosmology, cosmic variance of the local $H_0$ 
has been estimated in \cite{Shi:1997aa,Wang:1997tp,Buchert:1999pq,Wojtak:2013gda}. First attempts 
to estimate cosmic variance of the local Hubble rate in a relativistic approach can be found in 
\cite{Li:2007ny, Wiegand:2011je} (see also \cite{ClarksonetAll}), based on the ensemble variance
of the expansion rate averaged over a spatial volume. It has been shown that this approach 
agrees very well with the Newtonian one \cite{Li:2007ny} and it predicts 
a cosmic variance which depends on the sampling volume
on the sub-per cent to per cent level. However, this approach still neglects the fact that
observers probe the past light-cone and not a spatial volume. Also, the measured quantity is not an 
expansion rate, but a set of the bolometric fluxes and redshifts.

In this letter, we present the first fully relativistic estimation of the effects of clustering on the 
local measurement of the Hubble parameter without making any special hypothesis about 
how the fluctuations can be modeled around us. 
Considering only the 
measured quantities and the cosmological standard model with
stochastic inhomogeneities, we study the effect of cosmic structures on the local 
determination of $H_0$, i.e., taking light propagation effects fully into account.
Other relativistic approaches were recently proposed in \cite{FDU,Marra:2013rba}.
In  \cite{FDU} a "Swiss cheese" model was used in modeling the local Universe, 
in \cite{Marra:2013rba} a ``Hubble bubble'' model was used and the perturbation of the expansion 
rate, which is not directly measurable, was considered.

We shall find that the mean value of the Hubble parameter is modified at sub-percent level, while the 
contribution from clustering to the error budget is larger, typically $2$ to $3\%$,
hence as large as observational errors quoted in the literature~\cite{Riess}.  
As we shall see, the small modification of the mean of the Hubble parameter
can be reduced by a factor of $3$ by using the flux 
instead of the distance modulus.
On the other hand, the cosmic variance induced by inhomogeneities on $H_0$ is independent of 
the observable used. Finally, we find that even for an infinite number of SNIa within $0.01<z<0.1$
with identical redshift distribution compared to a finite sample considered, 
clustering induces a minimal error of about $2\%$ for a local determination of $H_0$.

Following \cite{BGMNVprl,BGMNVfeb2013} 
we use cosmological perturbation theory up to second order with an almost scale-invariant initial 
power spectrum to determine the mean perturbation of the bolometric flux (and of the distance modulus) 
from a standard candle and its variance. 

Let us first consider the fluctuation of the mean on a sphere at fixed observed redshift $z$. 
We denote the light-cone average \cite{GMNV} over a surface at fixed redshift by $\langle \cdots\rangle$,  
and a statistical average by $\overline{\cdots}$. Using the results of \cite{BMNG,FGMV} (see also 
\cite{BDG}) the fluctuation of the flux $\Phi \propto d_L^{-2}$, away from its background value in the 
Friedmann-Lema\^\i tre Universe (denoted by $ (d_L^{\rm FL})^{-2}$), is given by 
\be
 d_L^{-2}= (d_L^{\rm FL})^{-2} \left[1 + \Phi_1/\Phi_0 + \Phi_2/\Phi_0 \right]\,,
\label{EqFlux0}
\ee
where we expand $\Phi=\Phi_0+\Phi_1+\Phi_2$ up to second order in perturbation theory.
The ensemble average of $\langle \Phi_1/\Phi_0 \rangle$ vanishes at first order, 
but not at second order and must be added to another second order contribution from $\Phi_2/\Phi_0$; 
we obtain (see, e.g.~\cite{BGMNV1})
\be
\overline{\langle d_L^{-2} \rangle}(z) = (d_L^{\rm FL})^{-2} \left[1 + f_{\Phi}(z) \right],
\label{EqFlux1}
\ee
where for $z \ll 1$
\be
 f_{\Phi}(z) \simeq -\left(\frac{1}{{\cal H}(z) \Delta \eta}\right)^2\overline{\langle\left(\vec{v}_s \cdot \vec{n}\right)^2\rangle}\, .
 \label{1}
\ee
Here $\vec{n}$ denotes the direction to a given SN and $\vec{v}_s$ its peculiar velocity,
$\eta$ is conformal time, $\De\eta=\eta_0-\eta(z)$ is the difference between the present time and the time at redshift $z$, 
and ${\cal H}$ is the conformal Hubble parameter.
In \cite{BGMNVfeb2013} the full contribution is given in terms of 39 Fourier integrals over the dimensionless power 
spectrum of the Bardeen potential today, ${\cal P}_\psi (k)= (k^3/2 \pi^2)|\Psi_k(\eta_0)|^2$  with different kernels. 
We have removed the observer velocity since observations are usually quoted in the CMB 
frame, corresponding to $\vec{v}_0=0$. A non-vanishing observer velocity would nearly 
double the effect in Eq.~(\ref{1}). The dominant peculiar velocity contribution at low redshift gives
\bea \hspace*{-0.4mm}
f_\Phi(z)\simeq -\left(\!\frac{1}{{\cal H}(z) \Delta \eta}\!\right)^2\!\!\frac{\tau^2(z)}{3}\!\!
 \int_{H_0}^{k_{\rm UV}}\!\!\! \frac{dk}{k} k^2 {\mathcal P}_\psi (k),
\label{Spacek}\eea
where 
$$ \tau(z) =\int_{\eta_{in}}^{\eta_{s}} d\eta \frac{a(\eta)}{a(\eta_{s})}  \frac{g(\eta)}{g(\eta_0)}\, .$$
$g(\eta)$ is the growth factor and the source and the observer times are indicated with the suffix $s$ and $0$.

The brightness of supernovae is typically expressed in terms of the distance modulus $\mu$. 
Due to the nonlinear function relating $\mu$ and $\Phi$ one obtains different second order contributions,
\be
\overline{\langle \mu \rangle}- \mu^{\rm FL}= - \frac{2.5}{ \ln(10)}
\left[ f_{\Phi}-\frac{1}{2}\overline{\langle \left(\Phi_1/\Phi_0\right)^2 \rangle}
\right] \, ,
\label{2}
\ee
where, at $z \ll 1$, we also find
\be
\overline{\langle \left(\Phi_1/\Phi_0\right)^2 \rangle} \simeq -4f_\Phi \, .
\label{OTsmallz}
\ee

The approximate equalities in Eqs.~(\ref{1}) and~(\ref{OTsmallz}) are valid for $z\ll1$, where the first 
order squared contribution of the peculiar velocity terms dominates over the other second order 
contributions. 
For $z\sim 0.3$ and larger, additional contributions notably due to lensing become relevant, 
see~\cite{BGMNVprl,BGMNVfeb2013}.

For measurements of the Hubble parameter, low redshift SNe are used in order to minimize the 
dependence of the result on cosmological parameters. 
As a consequence, Eqs. (\ref{1}) and~(\ref{OTsmallz}) are good approximations for the aim of this Letter.

Hereafter we use the cosmological parameters from Planck~\cite{Planck}, the linear transfer function given in 
\cite{EH} taking baryons into account, and $k_{\rm UV}=0.1\,h\, \rm{Mpc}^{-1}$, see \cite{BGMNVfeb2013} 
for details. Increasing the cut-off does not change our result due to two effects: 
the kernel $k^2 {\mathcal P}_\psi (k)$ of the peculiar velocity contribution
decreases at large $k$ and small scale fluctuations are incoherent (see below) and their contribution 
to the variance decays like $1/N$, where $N$ is the number of supernovae.

\begin{figure}
\centering
\includegraphics[width=8cm]{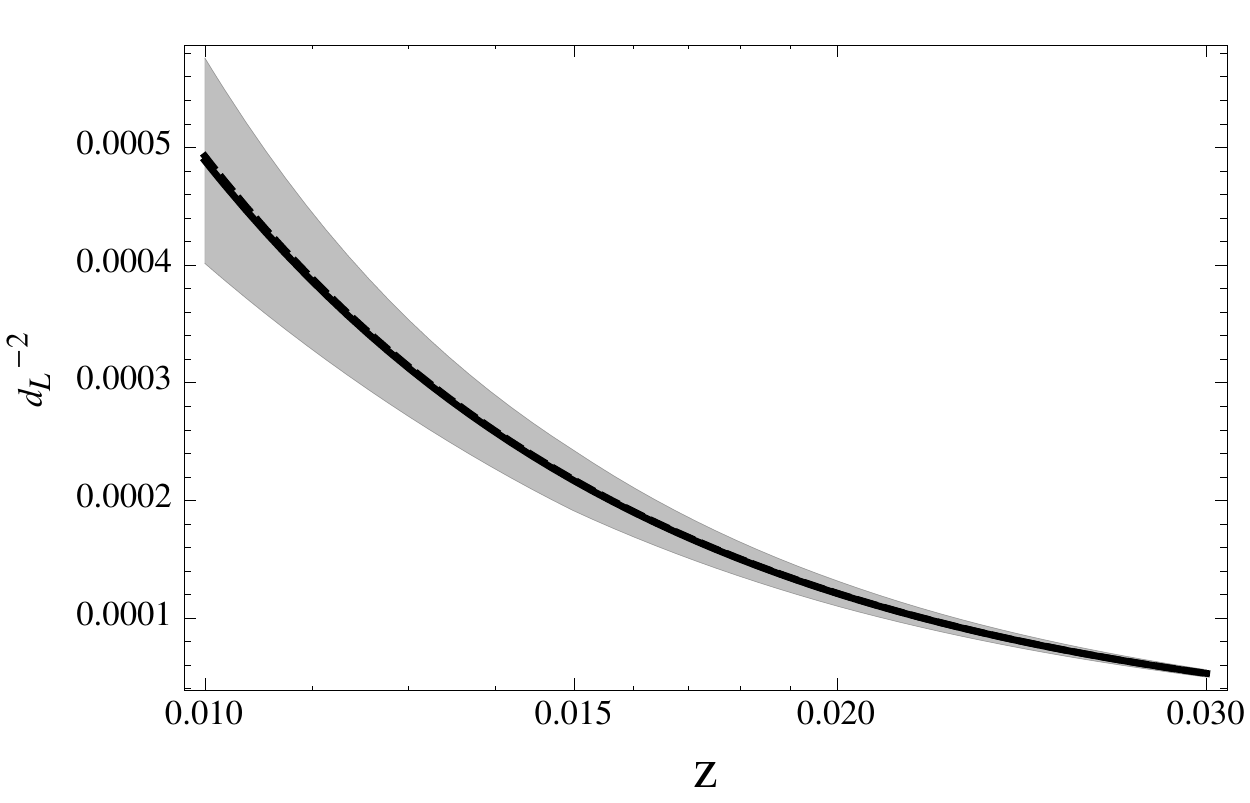}\\
\centering
\caption{The average  $\overline{\langle d_L^{-2} \rangle}(z)$ of Eq. (\ref{1}) in units of Mpc$^{-2}$ 
(thick solid curve), its dispersion (shaded region), and the homogeneous value (dashed curve) are 
computed within a range $z=0.01$ and $z=0.03$. We have used a best-fit cosmology from 
Planck~\cite{Planck} and a UV cut-off of $k_{\rm UV}=0.1\,h\, \rm{Mpc}^{-1}$.} 
\label{Fig1}
\end{figure}

As an illustration for the effects of cosmic structure on the observed flux from SN, we plot in 
Fig.~\ref{Fig1} the average $\overline{\langle d_L^{-2} \rangle}(z)$ and its variance (as defined in \cite
{BGMNV1}), using Eqs.~(\ref{EqFlux1}-\ref{Spacek}) and (\ref{OTsmallz}).
Figure \ref{Fig1} clearly shows how at low redshift 
the dispersion of the flux is much more important than the shift of the average (see also \cite
{BGMNVprl,BGMNVfeb2013}).

Comparing Eqs.~(\ref{EqFlux1}) and (\ref{2}), 
we see that the flux averaged over a sphere at constant redshift,  experiences a different effect than 
the distance modulus averaged over the same sphere.  

On the other hand, the induced theoretical dispersion on the bare value of $H_0$, which is entirely due 
to squared first order perturbations, is independent of the observable considered to infer $H_0$.
To determine the dispersion of $H_0$ from a sample of SNe we consider that at 
small redshift $H_0^2 \simeq c^2 z^2/d_L^2$. $H_0$ inferred from the observation  
of a single SN at redshift $z\ll 1$, is then expected to deviate from the true $H_0$ by approximately
\cite{BGMNV1}
\be
(\De H_0)^2 =\frac{H_0^2}{4}\overline{\langle \left(\Phi_1/\Phi_0\right)^2 \rangle} \label{e:auto}\,.
\ee
Of course in practice, observers do not have at their disposal many SNe at the same redshift, so the 
average over a sphere cannot be performed. Hence, we now go beyond this simplifying 
assumption of previous works. 

Let us estimate the (ensemble) variance of the locally measured 
Hubble parameter $H_0$ from the covariance matrix of the 
fluxes, given an arbitrarily distributed sample of $N$ observed SNe at 
positions $(z_i,\vec{n}_i)$, which reads 
\bea
\left(\frac{\De H_0}{H_0}\right)^2 &=&\frac{1}{4N^2}\sum_{ij} 
\overline{ 
\frac{\Phi_1(z_i,\vec{n}_i)}{\Phi_0(z_i)}\frac{\Phi_1(z_j,\vec{n}_j)}{\Phi_0(z_j)}}  \nonumber\\
& =& \frac{1}{N^2}\sum_{ij} \frac{V_{ij}}{\HH(z_i)\De\eta_i\HH(z_j)\De\eta_j}\, ,
\label{e:DeH0}
\eea
with
\be
\!V_{ij} = \!\tau(z_i)\tau(z_j)\int_{H_0}^{k_{\rm UV}}\!\frac{dk}{k}k^2 {\mathcal P}_\psi (k)I\big(k\De\eta_j,k\De\eta_i, (\vec{n}_i\cdot\vec{n}_j)\big),
\ee
and
\bea
I(x,y,\nu) &=&\frac{1}{4\pi}\int d\Om_{\hat k}
e^{ix(\hat{k}\cdot\vec{n}_i)}
e^{-iy(\hat{k}\cdot\vec{n}_j)}(\hat{k}\cdot\vec{n}_j)(\hat{k}\cdot\vec{n}_i)
\nonumber\\
&=&\frac{xy(1\!-\!\nu^2)}{R^2}j_2(R)\!  +\! \frac{\nu}{3}\big[j_0(R)-2j_2(R) \big],  
 \label{e:fI}
\eea
where $\nu= (\vec{n}_i\cdot\vec{n}_j)$ and $R=\sqrt{x^2+y^2-2\nu xy}=kd$.
Here $d$ is the comoving distance between the SNe at $(z_i,\vec{n}_i)$ and $(z_j,\vec{n}_j)$, $j_\ell$ denotes 
the spherical Bessel function of order $\ell$ and $\hat k$ is the unit vector in direction $\vec k$.
To arrive at (\ref{e:fI}), we have introduced the Fourier representation of 
$\Phi_1(z_i,\vec{n}_i) = 2/(\HH(z_i)\De\eta_i) \vec{v}_s(\vec{k})\cdot\vec{n}_i$ and 
used some well known identities.
Note that  with $I(x,x,1)=1/3$ and Eqs.(\ref{Spacek}) and (\ref{OTsmallz}), the auto-correlation term reproduces (\ref{e:auto}).

If the fluxes are perfectly coherent for all SNe so that $\overline{\Phi_1(z_i,\vec{n}_i)\Phi_1(z_j,\vec{n}_j)}= 4\si^2\Phi_0(z_j)\Phi_0(z_i)$, for all correlations, we obtain
$(\De H_0/H_0)^2 =\si^2$, while in the incoherent case, $\overline{\Phi_1(z_i,\vec{n}_i)\Phi_1(z_j,\vec{n}_j)} = \de_{ij}4\si^2\Phi_0(z_j)\Phi_0(z_i)$ we obtain
$(\De H_0/H_0)^2 =\si^2/N$. The reality lies somewhere in-between, wavelengths with $k d <1$ being rather coherent while those with $k d >1$ are rather incoherent.

\begin{figure}
\centering
\includegraphics[width=8cm]{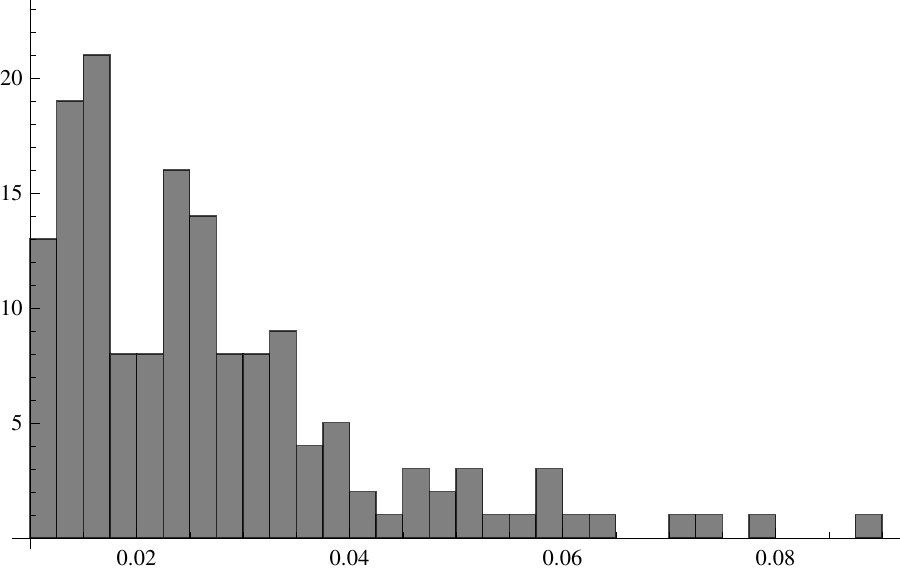}\\
\centering
\caption{The redshift distribution of the 155 SNe of the CfA3+OLD sample \cite{Hicken,Jha:2006fm} with redshift within $0.01$ and $0.1$ considered here.}
\label{fig:zdist}
\end{figure}

In order to estimate the effect of the cosmic (co)variance for a realistic sample of SNe, we consider 
the following set up. 
We calculate $\De H_0/H_0$ from Eqs.~(\ref{e:DeH0}) to (\ref{e:fI}) considering 
the redshifts of a sample of 155 SNe selected to lie in the range $0.01\leq z\leq 0.1$ 
from the CfA3 and OLD samples  \cite{Hicken,Jha:2006fm}.
The redshift distribution of the sample is shown in Fig~\ref{fig:zdist}.
We do not use their actual positions on the sky (see below).
We then also study the limiting case of infinitely many SNe.

For the redshift distribution of the 155 SNe of this sample, Eq.~(\ref{e:DeH0}) yields a 
dispersion induced by inhomogeneities between $2.2$ and $3.3\%$ 
for different angular distributions for the SNe. From this range we infer
\be 
 \Delta H_0= (1.6 \div 2.4)\,\, {\rm km~s}^{- 1} {\rm Mpc}^{-1} \,,
\label{ErrorTheoretical}
\ee
with $H_0$ as given in~\cite{Riess} (where "...$\div$..." stands for "from...to..."). 
We have kept  $\nu$ constant to different values and we 
have chosen a random distribution of directions over one hemisphere. The different choices give rise to 
the range quoted above. The smallest error corresponds to a random distribution 
of directions over one hemisphere, while the largest one corresponds to the case where all SNe are 
inside a narrow cone ($\nu\simeq 1$). The dispersion due to the actual angular distribution of real SN samples
is left for future studies.

Let us also estimate the effect of inhomogeneities on the measured value of $H_0$ itself for 
this sample. In \cite{Riess} a partial reconstruction of the peculiar velocity field has been applied,
which however comes from the density field in the neighborhood of the SNe and 
therefore contributes only an incoherent part which we neglect.
Considering a perfectly homogeneous Universe, a measured Hubble parameter $\hat{H}_0$ is deduced 
from the measurement of 
$\mu (z\ll1) \simeq 5 \log (c z/\hat{H}_0)+C,$ with $C$ a constant, 
see Eqs.~(\ref{mudef},\ref{dLLCDM}). 
However this is not the true underlying $H_0$, since it ignores the local large-scale structure, and therefore gives a biased value.
The true underlying Hubble parameter is derived only by applying the appropriate correction due to this structure. Comparing Eq.~(\ref{2}) with the above expression, we have:
  \be 
H_0 \simeq \hat{H}_0 \!
\left(1-\frac{3}{2} f_\Phi\right).
\label{RiessVSbackground}
\ee

We now consider the 155 SNe of the sample used here and generate the 
mean value of the corrected $H_0$, starting from a value of $\hat{H}_0$
and for the given redshift distribution. The final result is about $0.3\%$ higher than 
$\hat{H}_0$\footnote{Choosing a larger cut-off affects only this result slightly.}. 
A similar global shift has already been included in the analysis of~\cite{Riess} as a consequence of the partial reconstruction of the peculiar velocity field \cite{Riess1}.
Let us underline that the correction to $H_0$ would be three times  smaller if we would consider 
the backreaction on the {\em flux} instead of the one on the {\em distance modulus}. In this case Eq.~(\ref{RiessVSbackground}) should be replaced by $H_0 \simeq \hat{H}_0 \left(1-\frac{1}{2} f_\Phi\right)$.

Considering the quoted observational error of 2.4 km/s/Mpc~\cite{Riess} and
the additional variance~(\ref{ErrorTheoretical}), we obtain
\begin{eqnarray}
H_0 &=& 
\left[ 73.8 \pm 2.4 \pm (1.6 \div 2.4) \right]\,\, {\rm km~s}^{- 1} {\rm Mpc}^{-1}. 
\label{FinalResult2} 
\end{eqnarray} 
The tension with the Planck measurement \cite{Planck}, for which a value 
$(H_0)_{\rm CMB }= 67.3 \pm 1.2\,\, {\rm km~s}^{- 1} {\rm Mpc}^{-1}$ is reported,  is reduced when taking 
this additional variance into account. In particular, adding the above errors in quadrature we obtain a 
deviation of $2.2$ to $1.9 \, \sigma$ from  
$(H_0)_{\rm CMB }$, while the difference  is $2.7 \sigma$ when using the error quoted in \cite{Riess}.
This analysis is insensitive to smaller scales fluctuations due to the incoherence of such contributions. 
Further modeling of these scales, e.g.~\cite{FDU} (see also \cite{Fleury:2014gha}), might increase the uncertainty. 
However, effects from nearby small-scale structure are at least partly included in the analysis of~\cite{Riess}.

Before concluding, we want to determine the ultimate error for an arbitrarily large sample when the 
SNe are distributed isotropically over directions. In this case we can integrate $I(x,y,\nu)$ over all  
directions. With 
$$
\frac{1}{2}\int_{-1}^1d\nu I(x,y,\nu) = j_1(x)j_1(y) 
$$
we obtain, for a normalized redshift distribution $s(z)$,
\be
\!\left(\!\frac{\De H_0}{H_0}\right)^2\!\!\!\!=\!\!\!\int\!\!\frac{dk}{k}k^2
{\cal P}_\psi(k) \!\!\left(\!\int\!\! dz\tau(z) s(z)
 \frac{j_1(k\De\eta(z))}{\HH(z)\De\eta(z)}\right)^2
\label{LastRes}
\ee
with $\int dz s(z)=1$.
Approximating the redshift distribution of our sample
using an interpolating function of the histogram in Fig~\ref{fig:zdist}, integrating from $z=0.01$ to $0.1$, we 
obtain a dispersion of about $1.8\%$ which corresponds to an error of 
\be\label{e:H0z0.1}
\Delta H_0= 1.3\,\, {\rm km~s}^{- 1} {\rm Mpc}^{-1} \,.
\ee
This is the minimal dispersion of a SN sample with a redshift space distribution given by the one in 
Fig~\ref{fig:zdist}. It is not much smaller than the value obtained for the real sample. 
Interestingly,  this result is close to the ones obtained in 
\cite{Li:2007ny, Marra:2013rba, Wojtak:2013gda}, some of them with a very different analysis.

The errors from the nearby SNe with small $\De\eta(z)$ give the largest contribution. Therefore, the dispersion 
can be reduced by considering higher redshift SNe for which, however, the model dependence becomes more 
relevant. If we consider higher redshifts (close to or larger than $0.3$), we have to take into account also the 
other contributions to the perturbation of the luminosity distance, see \cite{BMNG,FGMV,BDG} for the full 
expression. As it is well known (see, for example, \cite{BGMNVprl,BGMNVfeb2013}), at redshift $z>0.3$, the 
lensing term begins to dominate.  

In~\cite{Neill} the peculiar velocity field has been reconstructed using the IRAS PSCz catalog~\cite{Branchini}. 
As already mentioned above, this is subtracted in the analysis of~\cite{Riess}. It is clear that 
this procedure also modifies the expected mean and its variance in our method, but a detailed analysis of 
this is beyond the scope of this work. As the (minimal) cosmic variance Eq.~(\ref{e:H0z0.1}) receives 
mainly contributions from scales larger than those considered in the reconstruction, we expect that it still has to 
be taken into account, in addition to the reconstructed peculiar velocities.

To conclude, in this Letter we  estimate the impact of stochastic inhomogeneities 
on the local value of the Hubble parameter and on its error budget for a given sample of 
standard candles. Eqs.~(\ref{e:DeH0}) to (\ref{e:fI}) and (\ref{LastRes}) are the main result of this Letter, namely 
a general formula for the cosmic variance contribution to $\De H_0$ from a sample of SNe with $z \lsim 0.2$, where the Doppler 
term dominates, and its limit for an arbitrarily large number of SNe isotropically distributed over directions.
This general formula can be easily implemented and does not require an N-body simulation for each set of cosmological parameters. The required input are solely the linear power spectrum and
the distribution of the observed SNe in position and redshift space.
In particular, we have found that for samples presently under consideration, this error 
is not negligible but of the same order as the experimental error, i.e.~between $2.2$ and $3.3\%$. 
We have also considered different samples (e.g.~95 SNe from \cite{Hicken}), in the range 
$0.01<z<0.1$, and found similar results.
This cosmic variance is a fundamental barrier on the precision of a local measurement of $H_0$. 
It has to be added to the observational uncertainties and it reduces the
tension with the CMB measurement of $H_0$~\cite{Planck}.

Finally, even when the number of SNe is arbitrarily large, an irreducible error remains due to 
cosmic variance of the local Universe. We have estimated this error and found it to be about $1.8\%$ for 
SNe with redshift $0.01<z<0.1$ and a distribution given by the one in Fig.2. This error can only be reduced 
by considering SNe with higher redshifts, but if too high redshifts are included the result 
becomes strongly dependent on other cosmological parameters like $\Om_m$ and curvature.

We wish to thank Ulrich Feindt, Benedict Kalus, Marek Kowalski, Martin Kunz, Lucas Macri, Adam Riess, Mickael Rigault, Marco Tucci, Gabriele Veneziano and Alexander Wiegand for helpful discussion.
IB-D is supported by the German Science Foundation (DFG) within the
Collaborative Research Center (CRC) 676 Particles, Strings and the Early Universe. RD acknowledges the Swiss National Science Foundation.
GM is supported by the Marie Curie IEF, Project NeBRiC - "Non-linear effects and backreaction in classical and quantum cosmology". 
 DJS thanks the 
Deutsche Forschungsgemeinschaft for support within the grant RTG 1620 ``Models of Gravity''.


\end{document}